\documentclass[reprint,
 amsmath,amssymb,
 aps,
]{revtex4-2}

\usepackage{graphicx}
\usepackage{dcolumn}
\usepackage{systeme}
\usepackage{bm}
\usepackage{xcolor}
\usepackage{wrapfig}

\begin{document}

\preprint{APS/123-QED}

\title{Relativistic transformations of quasi-monochromatic optical beams}
\author{Murat Yessenov}
\author{Ayman F. Abouraddy}%
\affiliation{CREOL, The College of Optics \& Photonics, University of Central Florida, Orlando, FL 32816, USA}%

\date{\today}

\begin{abstract}
A monochromatic plane wave recorded by an observer moving with respect to the source undergoes a Doppler shift and spatial aberration. We investigate here the transformation undergone by a generic, paraxial, spectrally coherent quasi-monochromatic optical \textit{beam} (of finite transverse width) when recorded by a moving detector. Because of the space-time coupling engendered by the Lorentz transformation, the monochromatic beam is converted into a propagation-invariant pulsed beam traveling at a group velocity equal to that of the relative motion, and which belongs to the recently studied family of `space-time wave packets'. We show that the predicted transformation from a quasi-monochromatic beam to a pulsed wave packet can be observed even at terrestrial speeds.
\end{abstract}

\maketitle

An observer moving with respect to an optical source emitting a monochromatic plane wave (MPW) records a Doppler-shifted MPW \cite{Einstein1905ADP,Bohm2015Book,Taylor92Book,JaksonBook}. What are the changes observed by a detector moving with respect to a source emitting instead a generic monochromatic optical \textit{beam} (i.e., a transversely localized field)? Previously tackled questions regarding relativistic transformations of optical fields have sometimes revealed surprising answers. For example, Terrell \cite{Terrell1959PRL} and Penrose \cite{Penrose1959apparent} showed that the length of an object in an image captured by an instantaneous shutter does \textit{not} depend on the observer's velocity, thus disabusing the physics community of the notion of a `visible' Lorentz contraction \cite{Weisskopf1960PhysTod}. Recently, it has been shown that angular-momentum carrying optical fields exhibit exotic effects in a frame moving orthogonally to the optical axis, including an optical analog of the relativistic spin Hall effect \cite{Bliokh2012PRL,Bliokh2013JOpt}, transverse orbital angular momentum and spatio-temporal vortices \cite{Bliokh2012PRA,Bliokh2021PRL}, and relativistic spin-orbit interactions \cite{Smirnova2018PRA}. 

We analyze here the transformation of a generic quasi-monochromatic beam when recorded by an observer moving with respect to the source along the beam axis. Previous studies of such a transformation have revealed several mathematical results, the central one of which is that a strictly \textit{monochromatic} beam is recorded as a finite-bandwidth \textit{pulsed} beam \cite{Longhi04OE, Saari04PRE,Kondakci18PRL,Saari20JPC}. The space-time coupling engendered by the Lorentz transformation yields a propagation-invariant wave packet whose group velocity is the relative velocity between source and detector \cite{Kondakci18PRL}. Remarkably, the observed wave packet is a realization of so-called `space-time wave packets' (STWPs) \cite{Kondakci16OE,Parker16OE,Wong17ACSP2,Efremidis17OL,Porras17OL,Yessenov19PRA}, which have been recently synthesized via spatio-temporal spectral-phase modulation \cite{Kondakci17NP,Kondakci19NC,Bhaduri20NP,Yessenov22AOP}. 

Here we examine Lorentz transformations of spectrally coherent quasi-monochromatic optical beams, and provide an physically intuitive picture that underpins their conversion into STWPs. In this picture, an angularly induced Doppler broadening leads to the formation of an STWP, and we emphasize the impact of the beam's spatial width on the induced spatio-temporal field structure. Crucially, by relaxing the ideal monochromaticity assumption, we find that the spectral linewidth of the source determines a minimum observer velocity for these effects to be detectable. We examine the potential for observing such effects at terrestrial speeds with currently available narrow-linewidth lasers.

\begin{figure}[b!]
    \centering
    \includegraphics[width=8.6cm]{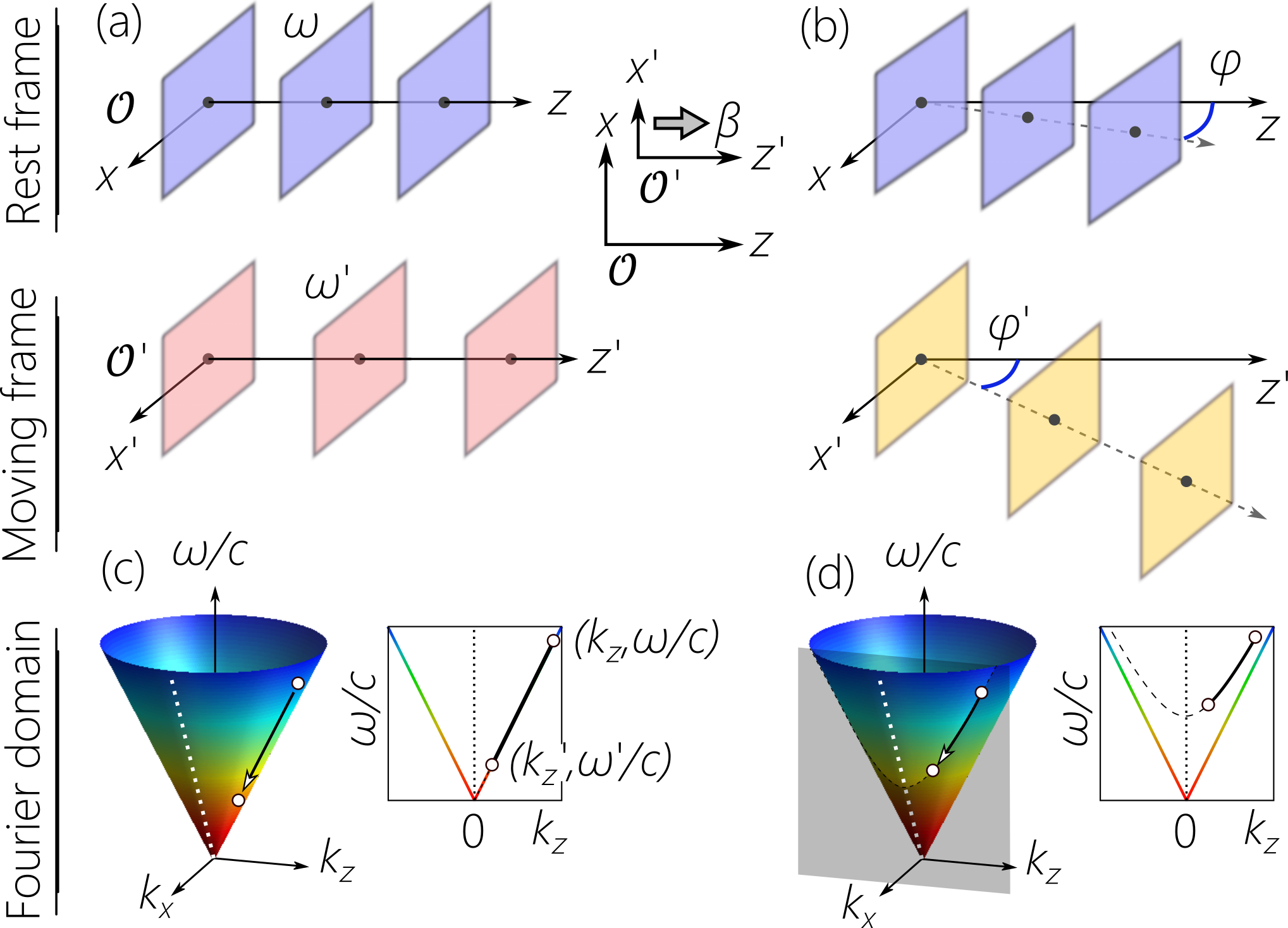}
    \caption{\small{A monochromatic plane wave (MPW) emitted in the rest frame $\mathcal{O}$ is Doppler-shifted in the frame $\mathcal{O}'$ moving along the $+z$-axis. (b) An off-axis MPW in $\mathcal{O}$ is Doppler-shifted and undergoes an angular rotation in $\mathcal{O}'$. (c) The on-axis MPW is Doppler-shifted along the light line $k_{z}\!=\!\omega/c$ ($k_{x}\!=\!0$) in the Fourier domain, whereas (d) an off-axis MPW is shifted along a fixed-$k_{x}$ hyperbola on the light-cone surface.}}
    \label{fig:Fig1}
\end{figure}

\begin{figure*}[t!]
    \centering
    \includegraphics[width=16.5cm]{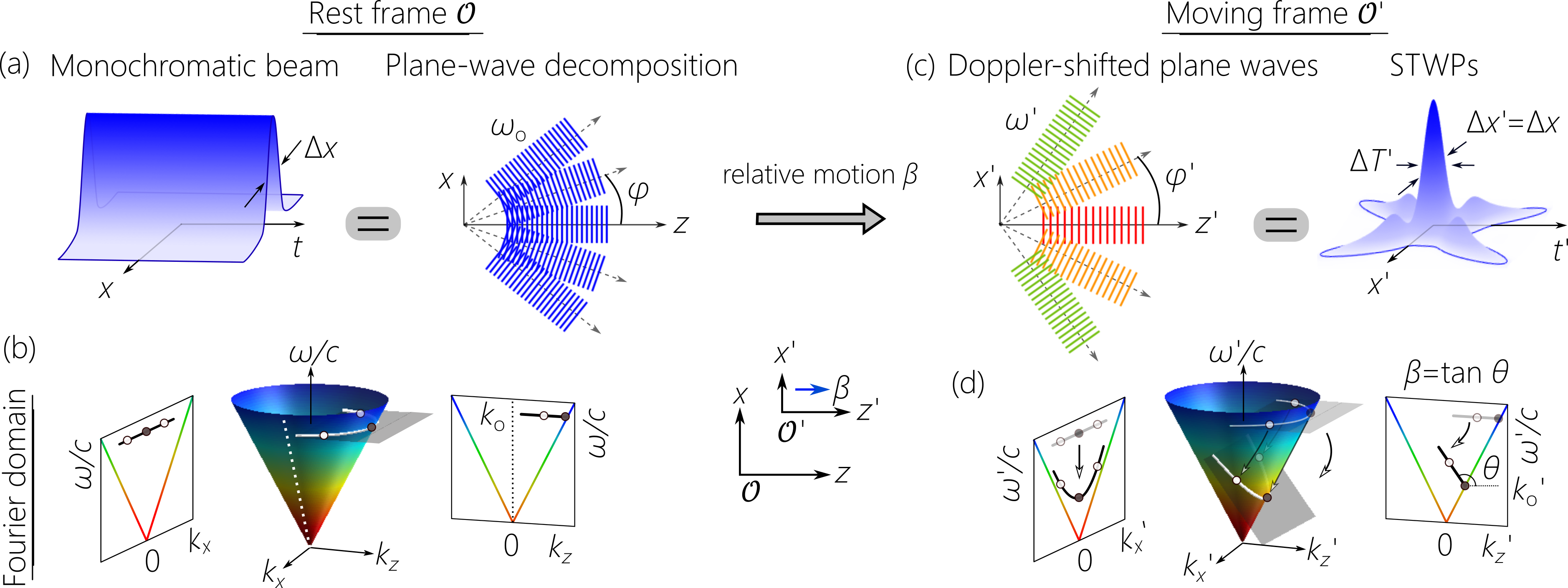}
    \caption{\small{Lorentz transformation of a monochromatic beam. (a) A monochromatic beam in $\mathcal{O}$ is a superposition of plane waves of the \textit{same} frequency $\omega_{\mathrm{o}}$ travelling in \textit{different} directions, and (b) its spectral support on the light-cone is an iso-frequency circle. (c) In the moving frame $\mathcal{O}'$, each plane wave undergoes an angle-dependent Doppler shift. (d) The spectral support for the field in (c) is the intersection of the light-cone with a plane that makes an angle $\theta$ with the $k_z'$-axis.}}
    \label{fig:Fig2}
\end{figure*}

To set the stage for analyzing the Lorentz transformation of optical beams, we first examine the case of MPWs in one transverse dimension $x$ (without loss of generality); see Fig.~\ref{fig:Fig1}. An MPW at frequency $\omega$ emitted by a source $\mathcal{S}$ at rest in the inertial frame $\mathcal{O}(x,z,t)$ is Doppler-shifted to $\omega'\!=\!\sqrt{\tfrac{1-\beta}{1+\beta}}\omega$ in the frame $\mathcal{O}'(x',z',t')$ moving at a velocity $v\!=\!\beta c$ along the common $z$-axis [Fig.~\ref{fig:Fig1}(a)]. An MPW travelling in $\mathcal{O}$ at an angle $\varphi$ with the $z$-axis is transformed in $\mathcal{O}'$ to a frequency $\omega'\!=\!\gamma(1-\beta\cos{\varphi})\omega$ travelling at an angle $\varphi'\!=\!\cos^{-1}{\left[\!\frac{\cos\varphi-\beta}{1-\beta\cos\varphi}\right]}$ (the Doppler spatial aberration \cite{Taylor92Book}), where $\gamma\!=\!1/\sqrt{1-\beta^{2}}$ [Fig.~\ref{fig:Fig1}(b)].

These changes can be visualized on the surface of the spectral light-cone \cite{Donnelly93ProcRSLA,Yessenov19PRA}. The wave vector $\vec{k}\!=\!(k_{x},k_{z})$ for an MPW in $\mathcal{O}$ is represented by a point on the surface $k_{x}^{2}+k_{z}^{2}\!=\!(\tfrac{\omega}{c})^{2}$, where $k_{x}\!=\!\tfrac{\omega}{c}\sin{\varphi}$ and $k_{z}\!=\!\tfrac{\omega}{c}\cos{\varphi}$. The Lorentz-transformed wave-vector components are: $k'_{x}\!=\!k_{x}; k_{z}'\!=\!\gamma(k_{z}-\beta\omega/c)$; and $\omega'\!=\gamma(\omega-c\beta k_{z})$. Because $k_{x}'^{2}+k_{z}'^{2}\!=\!(\tfrac{\omega'}{c})^{2}$, the structure of the light-cone itself is Lorentz invariant, so that the points corresponding to MPWs in $\mathcal{O}$ and $\mathcal{O}'$ can be represented on the same surface. The MPW in Fig.~\ref{fig:Fig1}(a) corresponds to a point on the light-line $k_{x}\!=\!0$, along which its Doppler-shifted counterpart in $\mathcal{O}'$ is displaced [Fig.~\ref{fig:Fig1}(c)]. In contrast, the point representing the off-axis MPW in $\mathcal{O}$ [Fig.~\ref{fig:Fig1}(b)] is displaced in $\mathcal{O}'$ along a constant-$k_{x}$ hyperbola [Fig.~\ref{fig:Fig1}(d)].

Now consider a generic monochromatic \textit{beam} emitted by the source $\mathcal{S}$ in $\mathcal{O}$ [Fig.~\ref{fig:Fig2}(a)], which is a superposition of plane waves (spatial bandwidth $\Delta k_{x}$, inverse of the beam width $\Delta x$) all at the same frequency $\omega_{\mathrm{o}}$ but traveling at different angles $\varphi$ with the $z$-axis \cite{GoodmanBook05,SalehBook07}. The spectral support for such a beam is the circle $k_{x}^{2}+k_{z}^{2}\!=\!k_{\mathrm{o}}^{2}$ at the intersection of the light-cone with a horizontal iso-frequency plane $\omega\!=\!\omega_{\mathrm{o}}$ [Fig.~\ref{fig:Fig2}(b)]; here $k_{\mathrm{o}}\!=\!\omega_{\mathrm{o}}/c$. Because the Doppler shift depends on the relative velocity $v$ and angle $\varphi$ between source and detector, the MPWs in $\mathcal{O}$ undergo \textit{different} Doppler shifts in $\mathcal{O}'$ [Fig.~\ref{fig:Fig2}(c)], and the associated points along the circle on the light-cone in $\mathcal{O}$ are displaced in $\mathcal{O}'$ differently along the constant-$k_{x}$ hyperbolas [Fig.~\ref{fig:Fig2}(d)]. Consequently, a finite spectral bandwidth $\Delta\omega'$ is Doppler-induced in the initially monochromatic beam, whose coherence guarantees that the space-time-coupled field in $\mathcal{O}'$ is \textit{pulsed} [Fig.~\ref{fig:Fig2}(c)]. The spectral support is Lorentz-transformed from a horizontal circle in $\mathcal{O}$ into a tilted ellipse in $\mathcal{O}'$ \cite{Kondakci18PRL} at the intersection of the light-cone with the plane $k_{z}'-k_{\mathrm{o}}'\!=\!(\omega'-\omega_{\mathrm{o}}')c\tan{\theta}$, which is parallel to the $k_{x}$-axis but makes an angle $\theta$ with the $k_{z}'$-axis, where $\tan{\theta}\!=\!-\beta$ [Fig.~\ref{fig:Fig2}(d)]. The linear relationship between $k_{z}'$ and $\omega'$ indicates the absence of dispersion in the observed wave packet, which travels in $\mathcal{O}'$ rigidly without diffraction at a group velocity $\widetilde{v}\!=\!c\tan{\theta}\!=\!-v$ \cite{Kondakci17NP,Kondakci19NC,Yessenov19PRA}. Such a field corresponds to a so-called subluminal `baseband' STWP \cite{Kondakci17NP,Yessenov19PRA,Yessenov22AOP}, which have been recently synthesized with group velocities in the range $0.07c\!<\widetilde{v}\!<\!c$ \cite{Kondakci19NC,Yessenov19OE,Hall21PRArecycling}. It will of course be challenging to produce such STWPs via relative motion between the source and detector.

\begin{figure}[b!]
    \centering
    \includegraphics[width=8.6cm]{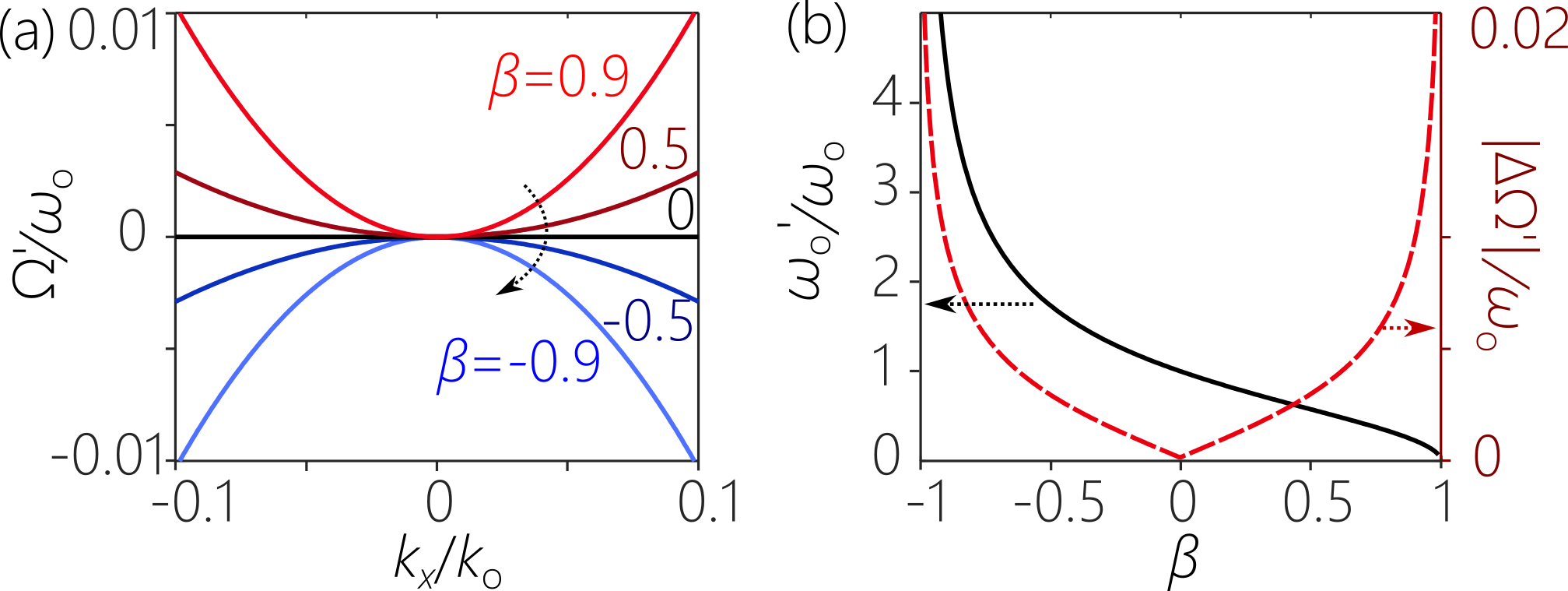}
    \caption{\small{(a) Spatio-temporal spectrum of paraxial STWPs in $\mathcal{O}'$ for different observer velocities $\beta$. (b) Dependence of the STWP central frequency $\omega_{\mathrm{o}}'$ (black solid curve, left axis) and bandwidth $|\Delta\Omega'|$ (red dashed curve, right axis) on $\beta$, normalized to the frequency $\omega_{\mathrm{o}}$ of the monochromatic beam in $\mathcal{O}$ having $\Delta k_{x}\!=\!0.1k_{\mathrm{o}}$.}}
    \label{fig:SpectralChanges}
\end{figure}

In the paraxial regime $\Delta k_{x}\!\ll\!k_{\mathrm{o}}$, the ellipse in $\mathcal{O}'$ can be approximated \cite{Kondakci17NP} by a parabola $\Omega'(k_{x}')\!=\!\frac{ck_{x}'^2}{2k_{\mathrm{o}}'(1-\cot{\theta})}$ [Fig.~\ref{fig:SpectralChanges}(a)], where $\Omega'\!=\!\omega'-\omega_{\mathrm{o}}'$. The initially monochromatic beam acquires a bandwidth $\Delta\Omega'\!=\!\tfrac{1}{2}\gamma|\beta|\omega_{\mathrm{o}}(\frac{\Delta k_{x}}{k_{\mathrm{o}}})^{2}$ via space-time coupling. Although $\Delta\Omega'$ is independent of the sign of $\beta$ (i.e., it is symmetric with respect to approaching or receding observers), the carrier frequency $\omega_{\mathrm{o}}'$ in contrast is highly asymmetric around $\beta\!=\!0$ [Fig.~\ref{fig:SpectralChanges}(b)]. The resulting on-axis ($x\!=\!0$) pulsewidth is $\Delta T'\!\sim\!\tfrac{1}{\gamma}\tfrac{z_{\mathrm{R}}}{v}$, where $z_{\mathrm{R}}$ is the Rayleigh range of the initial monochromatic beam. At a wavelength $\lambda_{\mathrm{o}}\!=\!800$~nm and beam width $\Delta x\!=\!40$~$\mu$m ($z_{\mathrm{R}}\!\sim\!1.6$~mm), relative motion at $v\!=\!0.8c$ results in $\Delta T'\!\sim\!4$~ps (a bandwidth $\Delta\lambda'\!\sim\!0.25$~nm). The pulsewidth is reduced to $\Delta T\!\sim\!250$~fs when the beam width is reduced to $\Delta x\!=\!10$~$\mu$m ($\Delta\lambda\!\sim\!4$~nm).

\begin{figure}[t!]
    \centering
    \includegraphics[width=8.4cm]{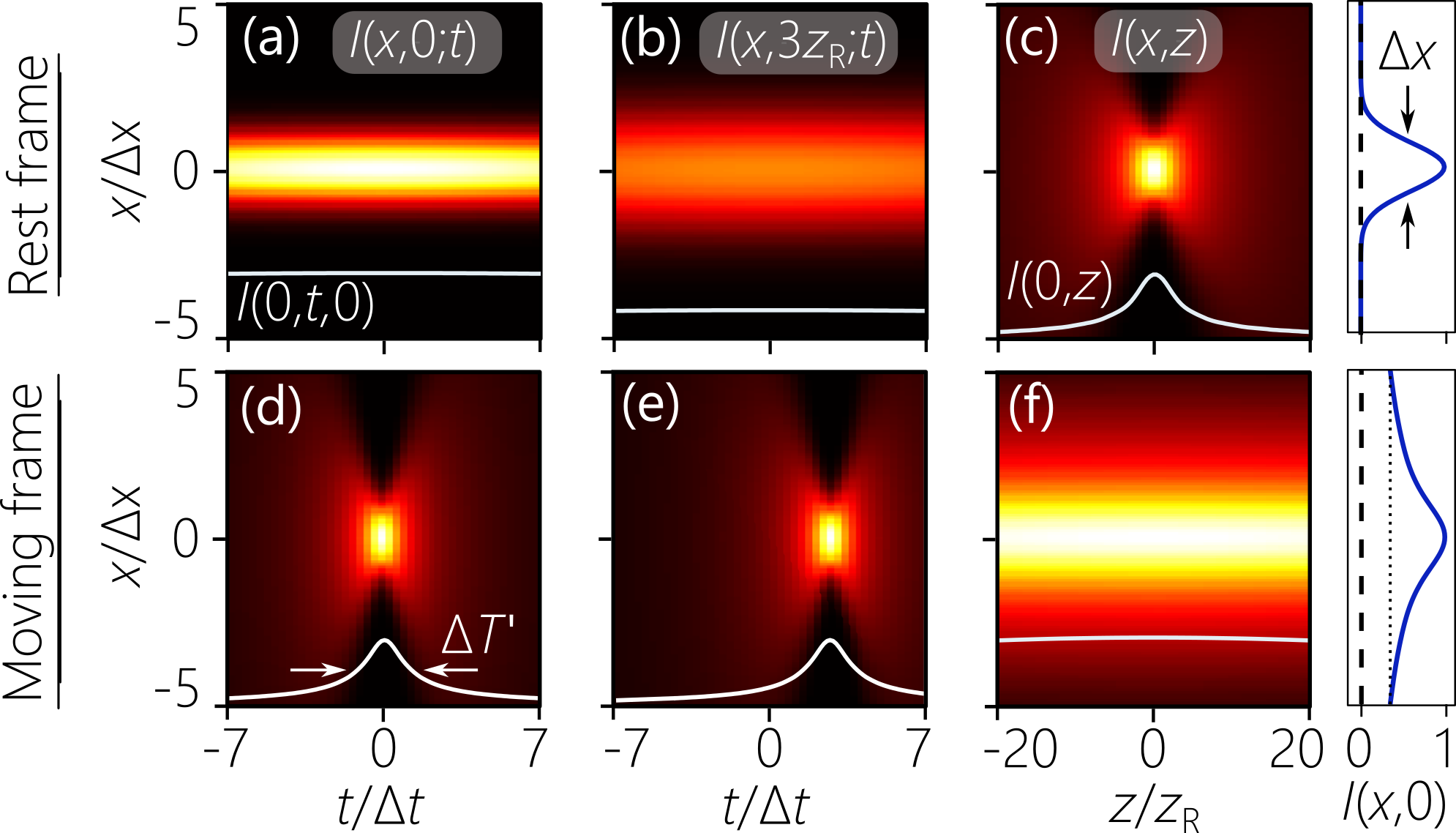}
    \caption{\small{(a) Spatio-temporal intensity profile $I(x,z;t)$ at the axial plane $z\!=\!0$ and (b) at $z\!=\!3z_{\mathrm{R}}$ in the rest frame $\mathcal{O}$. (c) The time-averaged intensity $I(x,z)$ in $\mathcal{O}$. (d) Spatio-temporal intensity profile $I(x',z';t')$ at $z'\!=\!0$ and (e) at $z'\!=\!3z_{\mathrm{R}}$ in $\mathcal{O}'$. (f) The time-averaged intensity $I(x',z')$ in $\mathcal{O}'$.}}\vspace{-5mm}
    \label{fig:SpatioTemporalChanges}
\end{figure}

Crucially, these conclusions are independent of the particular beam structure. In general, for a paraxial field $E(x,z;t)\!=\!e^{i(k_{\mathrm{o}}z-\omega_{\mathrm{o}}t)}\psi(x,z;t)$, the envelope is $\psi(x,z;t)\!=\!\iint\!dk_{x}d\Omega\,\widetilde{\psi}(k_{x},\omega)e^{i(k_{x}x+(k_{z}-k_{\mathrm{o}})z-\Omega t)}$. As a generic example, the spatio-temporal spectrum of a monochromatic Gaussian beam at $\omega_{\mathrm{o}}$ in the stationary frame $\mathcal{O}$ is $\widetilde{\psi}(k_{x},\omega)\!=\!\widetilde{\psi}(k_{x})\delta(\omega-\omega_{\mathrm{o}})$, with $\widetilde{\psi}(k_{x})\!\propto\!\exp\{-\tfrac{k_{x}^{2}}{2(\Delta k_{x})^{2}}\}$, so that $\psi(x,z;t)\!=\!\psi(x,z;0)$. The time-resolved intensity $I(x,z;t)\!=\!|E(x,z;t)|^{2}$ at any axial plane $z$ is thus independent of time [Fig.~\ref{fig:SpatioTemporalChanges}(a,b)]. Consequently, a `fast' detector recording $I(x,z;t)$ or a `slow' detector capturing the time-averaged intensity $I(x,z)\!=\!\int\!dt\,I(x,z;t)$ both reveal the \textit{same} spatial Gaussian envelope in $\mathcal{O}$ [Fig.~\ref{fig:SpatioTemporalChanges}(c)].

In the moving frame $\mathcal{O}'$, the spectrum is transformed into $\widetilde{\psi}(k_{x}',\omega')\!=\!\widetilde{\psi}(k_{x}')\delta(\Omega'-\Omega'(k_{x}'))$, from which it is straightforward to show that the envelope is propagation invariant $\psi(x',z';t')\!=\!\psi(x',0;t'-z'/\widetilde{v})\!=\!\psi(x',z'-\widetilde{v}t';0)$, with $\widetilde{v}\!=\!c\tan{\theta}\!=\!-v$. In other words, the roles of time and the axial coordinate $z'/\widetilde{v}$ are interchanged: the spatial profile observed along $z$ for a monochromatic beam in $\mathcal{O}$ [Fig.~\ref{fig:SpatioTemporalChanges}(c)] is now observed in time at a fixed axial plane in $\mathcal{O}'$ [Fig.~\ref{fig:SpatioTemporalChanges}(d,e)]. This phenomenon was predicted in \cite{Longhi04OE} where it was called `diffraction in time' (which is distinct from `time-diffraction' \cite{Moshinsky52PR,Moshinsky76AJP,Godoy02PRA}) and verified experimentally in \cite{Kondakci18PRL}. The invariance of $k_{x}$ in frames moving along $z$ guarantees that $\Delta x$ remains invariant. The time-averaged intensity $I(x',z')\!=\!I_{\mathrm{o}}+I_{\mathrm{ST}}(x')$ \cite{Kondakci19OL} is now independent of $z'$ and takes the form of a constant pedestal $I_{\mathrm{o}}$ atop of which is a Gaussian profile $I_{\mathrm{ST}}(x')\!=\!\int\!dk_{x}'|\widetilde{\psi}(k_{x}')|^{2}e^{i2k_{x}'x'}$ [Fig.~\ref{fig:SpatioTemporalChanges}(f)].

\begin{figure*}
    \centering
    \includegraphics[width=17.6cm]{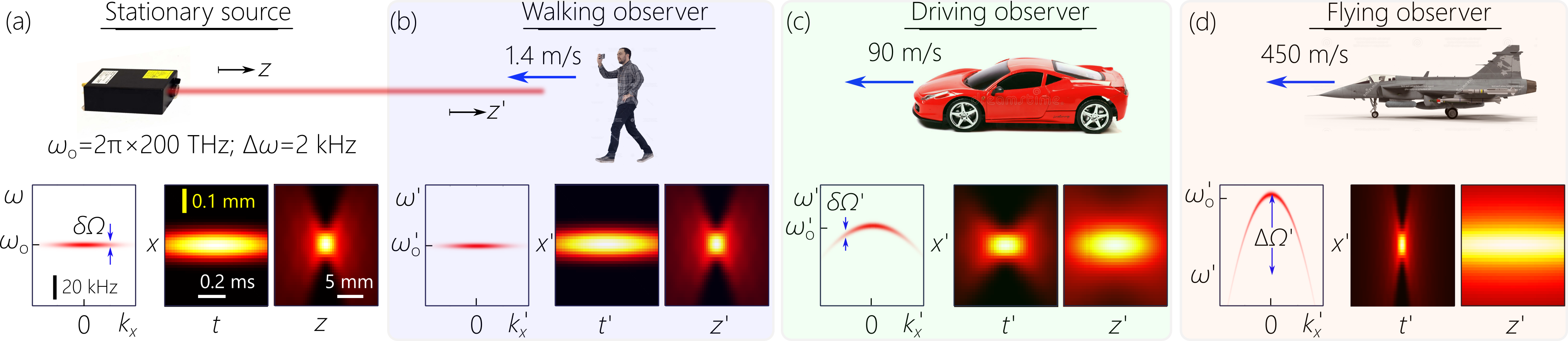}
    \caption{\small{Schematic of a potential test of relativistic transformations of a quasi-monochromatic beam. (a) A beam from a stationary laser ($\Delta x\!=\!100$~$\mu$m, $\tfrac{\omega_{\mathrm{o}}}{2\pi}\!=\!200$~THz, $\tfrac{\delta\Omega}{2\pi}\!=\!300$~Hz, and $z_{\mathrm{R}}\!=\!5$~mm) is recorded by moving observers. (b) A walking observer at $v\!=\!-1.4$~m/s ($\approx\!5$~km/h) does \textit{not} detect any change in the beam ($|v|\!<\!v_{\mathrm{min}}\!=\!40$~m/s). (c) A faster observer at $v\!=\!-90$~m/s ($\approx\!320$~km/h) detects a propagation-invariant STWP of pulsewidth $\Delta T'\!\approx\!0.5$~ms traveling at a group velocity $\widetilde{v}\!=\!90$~m/s, having a spatio-temporal spectral structure $\Omega'\!=\!\Omega'(k_{x}')$. (d) An even faster observer at $v\!=\!-450$~m/s ($\approx\!1600$~km/h) observes an STWP of pulsewidth $\Delta T'\!\approx\!0.1$~ms.}}
    \label{fig:SpectralUncert}
\end{figure*}

Recently, STWPs have been synthesized in the laboratory with $\widetilde{v}\!\sim\!c$ starting with generic pulsed beams \cite{Yessenov22AOP}. We inquire here whether relative motion at terrestrial velocities $v\!\ll\!c$ between a quasi-monochromatic source and a detector can lead to the observation of ultra-slow STWPs. To investigate this possibility, we must first drop the assumption of a strictly monochromatic field, for which any non-zero relative velocity can in principle lead to the formation of a detectable STWP via idealized space-time coupling $\delta(\Omega'-\Omega'(k_{x}'))$. Rather, a realistic finite-energy field is inevitably quasi-monochromatic, and hence possesses a finite linewidth $\delta\Omega$ in $\mathcal{O}$. When transformed in $\mathcal{O}'$ into an STWP, the precise delta-function correlation $\delta(\Omega'\!-\!\Omega'(k_{x}'))$ is relaxed to $g(\Omega'\!-\!\Omega'(k_{x}'))$ \cite{Yessenov19OE,Kondakci19OL}, where $g(\cdot)$ is a narrow spectral function whose width corresponds to a finite \textit{spectral uncertainty} $\delta\Omega'\!=\!\gamma(1-\beta)\delta\Omega$. The spectral uncertainty $\delta\Omega$ sets a minimum relative velocity $v_{\mathrm{min}}$ between source and detector that is required for a detectable STWP:
\begin{equation}\label{Eq:MinBeta}
v_{\mathrm{min}}\sim2c\left(\frac{\delta\Omega}{\omega_{\mathrm{o}}}\right)\bigg/\left(\frac{\Delta k_x}{k_{\mathrm{o}}}\right)^{2}=k_{o}\frac{(\Delta x)^{2}}{\delta T}\sim\frac{z_{\mathrm{R}}}{\delta T},
\end{equation}
where $\delta T\!\sim\!1/\delta\Omega$ is the pulsewidth of the field in $\mathcal{O}$.

This minimal requirement on the relative velocity can be understood from several perspectives. The spectral uncertainty $\delta\Omega$ is the finite bandwidth of the spectral support for the quasi-monochromatic field on the light-cone surface [Fig.~\ref{fig:Fig2}(b)]. The Doppler-induced bandwidth $\Delta\Omega'$ results in an on-axis pulsewidth $\Delta T'\!\sim\!\tfrac{1}{\Delta\Omega'}$ that is independent of the initial linewidth $\delta\Omega'$. For the relative motion to produce a detectable STWP, the spectral tilt angle $\theta$ must be sufficient for the new spectral support on the light-cone to be distinguishable from the initial spectrum. This requires that $\Delta\Omega'$ exceed the spectral uncertainty, $\Delta\Omega'\!>\!\delta\Omega'$, which sets a minimal spectral tilt angle, and hence a minimal relative velocity. A different perspective is gleaned from consideration of the maximum propagation distance of an STWP $L_{\mathrm{max}}\!\sim\!\tfrac{c}{\delta\Omega'|1-\cot{\theta}|}$ \cite{Yessenov19OE,Kondakci19OL}. Observing the STWP in $\mathcal{O}'$ requires that $L_{\mathrm{max}}$ be larger than the axial pulse length $v\Delta T'\!=\tfrac{z_{\mathrm{R}}}{\gamma}$, thereby leading to the result in Eq.~\ref{Eq:MinBeta}.

We illustrate  in Fig.~\ref{fig:SpectralUncert} the consequences of Eq.~\ref{Eq:MinBeta} starting with a quasi-monochromatic beam of $\tfrac{1}{e}$-width $\Delta x\!=\!100~\mathrm{\mu m}$ (Rayleigh range $z_{\mathrm{R}}\!\approx\!5$~mm) and spectral linewidth $\tfrac{\delta\Omega}{2\pi}\!=\!300$~Hz ($\delta T\!=\!1$~ms in $\mathcal{O}$) centered at $\tfrac{\omega_{\mathrm{o}}}{2\pi}\!=\!200$~THz ($\lambda_{\mathrm{o}}\!\approx\!1.55~\mathrm{\mu m}$) [Fig.~\ref{fig:SpectralUncert}(a)], which is observed by a moving detector [Fig.~\ref{fig:SpectralUncert}(b-e)]. From Eq.~\ref{Eq:MinBeta}, $\beta_{\mathrm{min}}\!=\!1.3\!\times\!10^{-7}$ or $v_{\mathrm{min}}\!\approx\!140$~km/h, so that an observer at $v\!=\!-5~\mathrm{km/h}$ ($|v|\!<\!v_{\mathrm{min}}$) records a conventionally diffracting quasi-monochromatic beam [Fig.~\ref{fig:SpectralUncert}(b)]. However, an observer at $v\!=\!-320~\mathrm{km/h}$ ($|v|\!>\!v_{\mathrm{min}}$) records an STWP with $\Delta T'\!\sim\!0.5$~ms,  $L_{\mathrm{max}}\!\approx\!10$~mm, and $\widetilde{v}\!=\!320$~km/h [Fig.~\ref{fig:SpectralUncert}(c)]. An even faster observer moving at $v\!=\!-1600$~km/h detects an STWP of shorter pulsewidth $\Delta T'\!\sim\!100$~$\mu$s and longer propagation distance of $L_{\mathrm{max}}\!\approx\!50$~mm [Fig.~\ref{fig:SpectralUncert}(d)].

Narrowing the linewidth to $\tfrac{\delta\Omega}{2\pi}\!=3$~Hz reduces the threshold to $v_{\mathrm{min}}\!\approx\!1.4$~km/h, and recording an STWP becomes accessible to a walking observer, whereas the flying observer records an STWP travelling freely for $L_{\mathrm{max}}\!=\!5$~m. Alternatively, $v_{\mathrm{min}}$ can be reduced more effectively by reducing the transverse beam width, due to the quadratic dependence $\beta_{\mathrm{min}}\!\propto\!(\Delta x)^{2}$. 

\begin{figure}[b!]
    \centering
    \includegraphics[width=8.6cm]{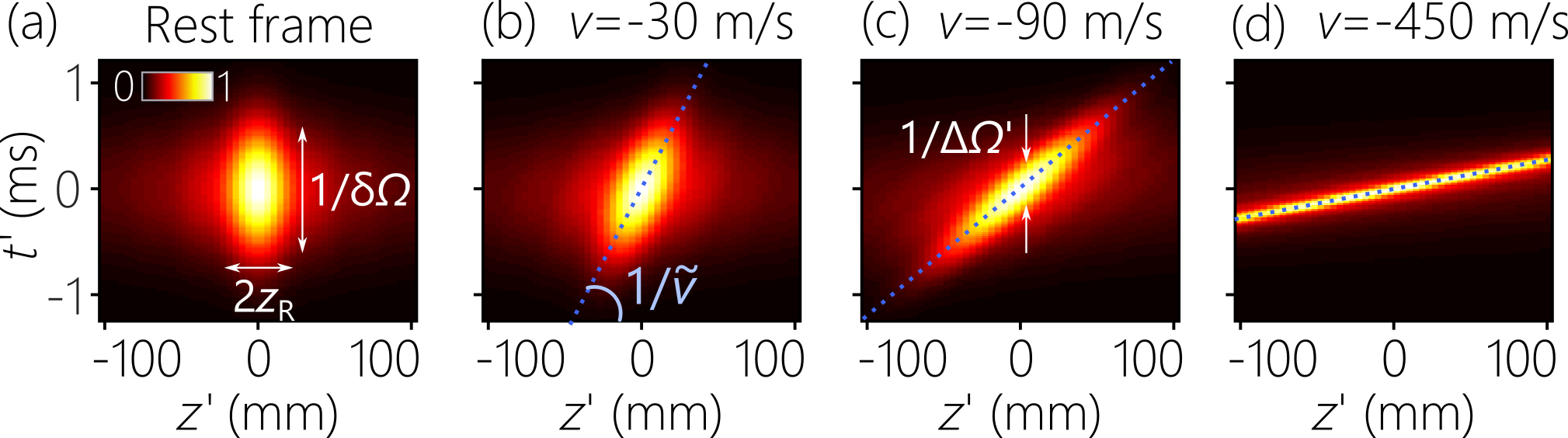}\vspace{-3mm}
    \caption{\small{On-axis intensity profile $I(x'\!=\!0,z';t')$ of a 100-$\mu$m-wide beam and 1-ms pulse duration observed in $\mathcal{O}'$ by (a) a stationary observer, or observers moving towards the source at (b) $v\!=\!-30$~m/s, (c) $-90$~m/s, and (d) $-450$~m/s.}}\vspace{-4mm}
    \label{fig:WorldLines}
\end{figure}

We plot in Fig.~\ref{fig:WorldLines} the on-axis intensity profiles $I(x'\!=\!0,z';t')$ with increasing $v$ at $\tfrac{\delta\Omega}{2\pi}\!=\!300$~Hz, which can be viewed as world-lines for the peak of the pulsed field in $\mathcal{O}'$. In $\mathcal{O}$, the long temporal extent of $\sim\!1$~ms (corresponding to a length of $\sim\!300$~km) combined with the short axial extent $\Delta z\!=\!2z_{\mathrm{R}}\!=\!10$~mm renders the \textit{peak} of the wave packet effectively `stationary', even though the underlying electromagnetic field is travelling at $c$ [Fig.~\ref{fig:WorldLines}(a)]. As the observer moves towards the source, the detected detected of the STWP can become significantly shorter than 1~ms when $v\!\gg\!v_{\mathrm{min}}$, resulting in an STWP peak moving at a group velocity $\widetilde{v}\!=\!-v$ [Fig.~\ref{fig:WorldLines}(b-d)], and with the STWP propagation invariant within the temporal interval of 1~ms [Fig.~\ref{fig:WorldLines}(d)]. 

We have considered here a model in which the initial laser spectrum is coherent. However, the narrow-linewidth spectra of realistic laser sources are largely \textit{incoherent}, corresponding to continuous-wave radiation rather than pulsed \cite{Henry82JQE,Pollnau20PQE}. The coherent spectral model utilized here can be obtained by modulating the narrow-linewidth source at a rate higher than its initial linewidth, resulting in a pulse train, each pulse of which is described by the model established here. The Lorentz transformation of a continuous-wave laser source with a spectrally incoherent narrow linewidth requires a different analysis \cite{Yessenov19Optica,Yessenov19OL}, which will be reported elsewhere. Furthermore, our analysis has been restricted to one transverse dimension, which has the advantage of showing a clear structure (the pedestal $I_{\mathrm{o}}$) emerging as a result of space-time coupling. Incorporating both transverse dimensions in an azimuthally symmetric beam does not change the conclusions except that the pedestal is replaced with a slow $\tfrac{1}{r}$-decay in intensity ($r$ is the radial coordinate) \cite{Yessenov2021Arxiv3D,Pang22OE}.


In contrast to previous tests of special relativity that rely on complex configurations and high-level precision \cite{Ives38JOSA,Hafele1972Science,Mansouri1977TestIII,Wolf2006Book,Chou10Science}, the conversion of a generic monochromatic paraxial beam into a propagation-invariant pulsed beam at terrestrial speeds is potentially simpler to realize, especially in light of the current availability of ultranarrow-linewidth lasers ($\tfrac{\delta\Omega}{2\pi}\!<\!300$~Hz) and high-speed cameras ($>\!1000$~frames/s). Although the Doppler shift is prohibitively difficult to detect at small $\beta$, the changes in the spatio-temporal structure of the field can be readily captured. Moreover, the results reported here may lead to new designs and functionalities for so-called space-time metasurfaces by elucidating what can be achieved at low-speed moving devices \cite{Leger2019AP,Caloz2020IEEE1,Caloz2020IEEE2,Rocca2020IEEE}. Other areas in optical physics have recently explored the ramifications of relativistic transformations of optical fields, including photonic time crystals \cite{Lyubarov2022Science,Sharabi2022Optica}; realizations of an optical analog of the Mackinnon wave packet \cite{Mackinnon78FP} via moving dipoles \cite{WilczekBook,Hall22NatPhys}; and reflection and refraction from moving surfaces \cite{Plansinis15PRL,Qu2016JOSAB,Plansinis16JOSAB,Plansinis17JOSAB,Plansinis18JOSAB,Leger2021Photonics}. Moreover, our findings may open a new perspective on transverse relativistic transformations of monochromatic beams carrying orbital angular momentum \cite{Bliokh2012PRA,Bliokh2012PRL,Bliokh2013JOpt, Smirnova2018PRA,Bliokh2021PRL}. 

In summary, we have analyzed a generic quasi-monochromatic optical beam observed in an axially moving frame, showing that the transformed field is a propagation-invariant wave packet of finite pulsewidth travelling at subluminal group velocities. Moreover, an intuitive physical picture provides the constraint on the relative velocity between source and detector required to observe the predicted phenomena. Our analysis reveals that current technology allows for such a test to be carried out at terrestrial speeds. 

\textit{Acknowledgments.} Authors thank P.~J.~Delfyett,  M.~G.~Vazimali and A. Duspayev for helpful discussions. This work was funded by the U.S. Office of Naval Research, contracts N00014-17-1-2458 and N00014-20-1-2789, and by the W. M. Keck Foundation. 

\clearpage

\bibliography{diffraction}


\end{document}


\preprint{APS/123-QED}

\title{Supplementary material for\\``Relativistic transformations of quasi-monochromatic optical beams''}
\author{Murat Yessenov}
\email{yessenov@knights.ucf.edu}
\author{Ayman F. Abouraddy}%
\affiliation{CREOL, The College of Optics \& Photonics, University of Central Florida, Orlando, FL 32816, USA}%

\date{\today}

\renewcommand{\thepage}{S\arabic{page}}  
\renewcommand{\thesection}{S\arabic{section}}   
\renewcommand{\thefigure}{S\arabic{figure}}

\maketitle

In this Supplementary document we provide details of the simulations to calculate the spatio-temporal profile of optical beams observed in frames moving with respect to the source (corresponding to Figs.~4-6 of the Main text).

\section*{Simulations of Lorentz transformation of a quasi-monochromatic optical field}

\subsection{Monochromatic beams}

We start our analysis from a monochromatic beam in the rest frame $\mathcal{O}(x,z,t)$ with a carrier frequency $\omega_{\mathrm{o}}$ and field profile $E(x,z;t)= e^{i(k_{\mathrm{o}}z-\omega_{\mathrm{o}}t)}\psi(x,z)$, where $\psi(x,z)$ is a slowly varying envelope, $x$ and $z$ are the transverse and axial coordinates, respectively, and $k_{\mathrm{o}}\!=\!\omega_{\mathrm{o}}/c$ is wave number associated with the carrier frequency \cite{SalehBook07}. The angular spectrum of the spatial envelope is given by:
\begin{equation}\label{Eq:SpatialEnvelope}
\psi(x,z)=\int dk_{x}\widetilde{\psi}(k_{x})e^{ik_{x}x}e^{i(k_{z}-k_{\mathrm{o}})z},
\end{equation}
where the spatial spectrum $\widetilde{\psi}(k_{x})\!=\!\int dx~\psi(x,0)e^{-ik_{x}x}$ is the Fourier transform of the initial profile $\psi(x,0)$. For example, for a Gaussian beam of transverse size $w_{\mathrm{o}}$ (1/e width), $\psi(x,z)$ correspond to the diffraction profile with a Rayleigh range $z_{\mathrm{R}}=k_{\mathrm{o}}w_{\mathrm{o}}^2/2$.

When such a monochromatic beam is observed in a different reference frame $\mathcal{O}'(x',z',t')$ in relative motion at a velocity $v\!=\!\beta c$  along a common $z$-axis, the field profile is transformed into a new form $E'(x',z';t')\!=\!E(x,z;t)$, with the relationship between the coordinates $(x',z',t')$ and $(x,z,t)$ determined by the Lorentz transformation:
\begin{eqnarray}
x&=&x',\nonumber\\
z&=&\gamma(z'+\beta ct'),\nonumber\\
t&=&\gamma(t'+\beta z'/c),
\end{eqnarray}
where $\gamma=1/\sqrt{1-\beta^2}$ is the relativistic Lorentz factor. The transformed field profile thus takes the form \cite{Belanger84JOSAA,Longhi04OE}:
\begin{eqnarray}\label{Eq:EPrimeMonochrom}
E'(x',z';t')&=&E(x',\gamma(z'+\beta c t');\gamma(t'-\beta z'/c))\nonumber\\ &=&e^{i(k_{\mathrm{o}}'z'-\omega_{\mathrm{o}}' t')}\psi(x',\gamma(z'+v t')), 
\end{eqnarray}
where $\omega_{\mathrm{o}}'=\gamma(1-\beta)\omega_{\mathrm{o}}$ is the Doppler-shifted carrier frequency, and $k_{\mathrm{o}}'=\omega_{\mathrm{o}}'/c$. It is clear from Eq.~\ref{Eq:EPrimeMonochrom} that the field in $\mathcal{O}'$ corresponds to a propagation-invariant pulsed beam travelling a group velocity $\widetilde{v}=-v$. The on-axis pulsewidth $\Delta T'$ of this propagation-invariant wave packet is dictated by the diffraction length of the monochromatic beam in the rest frame $\mathcal{O}$: $\Delta T'\!=\!\frac{\!z_{\mathrm{R}}}{\gamma v}$. Moreover, the transformed field is endowed with a one-to-one association between the spatial and temporal frequencies:
\begin{equation}\label{Eq:STCorrelation}
\omega'-\omega_{\mathrm{o}}'=(k_{z}'-k_{\mathrm{o}}')\widetilde{v},
\end{equation}
where $k_{x}'\!=\!k_{x}$, $k_{z}'\!=\!\gamma(k_{z}-\beta k_{\mathrm{o}})$, and $\omega'\!=\!\gamma(\omega_{\mathrm{o}}-\beta ck_{z})$ are the Lorentz-transformed spectral variables in $\mathcal{O}'$, and $k_{z}'\!=\!\sqrt{(\tfrac{\omega'}{c})^{2}-k_{x}'^{2}}$. This relationship is identical to the dispersion relationship characteristic of space-time wave packets (STWPs) \cite{Kondakci17NP,Yessenov19OPN,Yessenov22AOP}.
Thus, a generic diffractive monochromatic beam in the rest frame $\mathcal{O}$ is transformed into a propagation-invariant (diffraction-free and dispersion-free) STWP in the moving frame $\mathcal{O}'$ with tight space-time correlation $\omega'\!=\!\omega'(k_{z}')$. 

\subsection{Quasi-monochromatic beams}

Any realistic source possesses a finite spectral line-width. This limitation sets a minimum speed of the observer to detect this transformation, as we show below. In contrast to the ideal monochromatic case, a spectrally coherent quasi-monochromatic beam of spectral bandwidth $\delta\Omega$ corresponds to a pulse of width $\delta T\!\sim\!\tfrac{1}{\delta\Omega}$ that travels at a group velocity equal to $c$ in free space, which entails temporal dynamics during propagation; here $\Omega\!=\!\omega-\omega_{\mathrm{o}}$. In the paraxial regime, the field can be separated into the product of a spatial diffraction envelope and a plane-wave pulse envelope:
\begin{equation}\label{Eq:ParaxialQuasiMonochromatic}
E(x,z;t)=e^{i(k_{\mathrm{o}}z-\omega_{\mathrm{o}} t)}\psi_{x}(x,z)\psi_{t}(t-z/c),
\end{equation}
where $\psi_{x}(x,z)$ represents the spatial dynamics (diffraction pattern) at the carrier frequency $\omega_{\mathrm{o}}$ as considered in the previous Section (Eq.~\ref{Eq:SpatialEnvelope}), $\psi_{t}(t)\!=\!\int\!d\Omega\widetilde{\psi}(\Omega)e^{-i\Omega t}$ is a plane-wave pulse representing the temporal dynamics. Following the same procedure as in Eq.~\ref{Eq:EPrimeMonochrom}, the transformed field in $\mathcal{O}'$ takes the form: 
\begin{eqnarray}\label{Eq:EPrimePulsed}
&E&'(x',z';t')=E(x',\gamma(z'+\beta ct');\gamma(t'-\beta z'/c))\nonumber\\\!=\!&e&^{i(k_{\mathrm{o}}'z'\!-\!\omega_{\mathrm{o}}'t')}\psi_{x}(x',\gamma(z'\!+\!vt'))\psi_{t}(\gamma(1\!-\!\beta)(t'\!-\!z'/c)). 
\end{eqnarray}
This result is identical to the formulation of STWPs with finite spectral uncertainty as studied in \cite{Yessenov19OE}. In contrast to Eq.~\ref{Eq:EPrimeMonochrom}, the transformed optical field in Eq.~\ref{Eq:EPrimePulsed} separates into two terms. The first term $\psi_{x}(x',\gamma(z'+vt'))$ corresponds to an ideal propagation-invariant STWP of pulsewidth $\Delta T'\!=\!\tfrac{\!z_{\mathrm{R}}}{\gamma v}$ traveling at a group velocity $\widetilde{v}=-v$ \cite{Yessenov19OE}; whereas the second term $\psi_{t}(\gamma(1-\beta)(t'-z'/c))$ represents a plane-wave pulse of pulsewidth $\delta T'\!=\!\tfrac{\delta T}{\gamma(1-\beta)}$ travelling at $c$, which we have termed the `pilot envelope'\cite{Yessenov19OE}. It is important to note that the Doppler-induced pulsewidth of the STWP $\Delta T'$ is independent of the width of the pilot envelope $\delta T'$.

The separation of the initially quasi-monochromatic field into two terms in Eq.~\ref{Eq:ParaxialQuasiMonochromatic} thus entails the separation of the Lorentz-transformed field in Eq.~\ref{Eq:EPrimePulsed} in turn into two terme. Physically, these two terms correspond to two distinct pulses of finite durations and different speeds. This speed difference leads ultimately to walk-off in their product after a certain propagation distance $L_{\mathrm{max}}$ that determines the STWP diffraction-free length \cite{Yessenov19OE}. In order to detect this STWP, the observer must move fast enough such that the STWP pulsewidth is shorter than that of the pilot envelope: $\Delta T'<\delta T'$, which yields the condition:
\begin{equation}
c\delta T>\frac{1-\beta}{|\beta|}z_{\mathrm{R}}.
\end{equation}
When this condition is satisfied, an STWP with finite propagation distance $L_{\mathrm{max}}=\frac{c\delta T'}{1+1/\beta}=\gamma c\delta T/\beta$ is observed. 

In our simulations, we consider a conventional pulsed beam with Gaussian profiles in both space and time, $\tfrac{\omega_{\mathrm{o}}}{2\pi}\!=\!200$~THz ($\lambda_{\mathrm{o}}\approx1550$~nm). In the rest frame $\mathcal{O}$, the spatio-temporal intensity profile of the pulsed beam is:
\begin{equation}
I(x,z;t)\propto\frac{w_{o}}{w(z)}\exp{\left[-\frac{2x^2}{w(z)^2}-\frac{2t^2}{\delta T^2}\right]},
\end{equation}
where $w(z)=w_{\mathrm{o}}\sqrt{1+(z/z_{\mathrm{R}})^2}$, $w_{\mathrm{o}}$ is beam size at $z=0$ (1/e width), $z_{\mathrm{R}}=k w_{\mathrm{o}}^2/2$ is the Rayleigh length, we define $\Delta x=2w_{\mathrm{o}}$, and $\delta T$ is the initial pulsewidth \cite{SalehBook07}. In the moving frame $\mathcal{O}'$, the transformation given in Eq.~\ref{Eq:EPrimePulsed} yields:
\begin{eqnarray}
&&I'(x',z';t')\propto\frac{w_{\mathrm{o}}}{w(\gamma[z'+vt'])}\nonumber\\&&\exp{\left[-\frac{2x'^2}{w^2(\gamma[z'+vt'])}-\frac{2(\gamma[1-\beta][t'-z'/c])^2}{\delta T^2}\right]}.
\end{eqnarray} 

Alternatively, one can carry out the calculations in the spectral domain $(k_{x},k_{z},\omega)$ rather than in the physical domain $(x,z;t)$. By starting with a spatio-temporal spectrum $\widetilde{\psi}(k_{x},\Omega)$ in the rest frame, we transform it in the moving frame following a similar procedure to that in Eq.~\ref{Eq:EPrimeMonochrom}, and the field is propagated along $z$ using the Fourier-transform split-step method \cite{Wartak2013Book,Yessenov2020PRLaccel}. This method is particularly useful for beams whose spatial profile do not possess a closed-form expression. Our calculations were performed in both the physical domain and the spectral domain, and the same results were obtained. 


\bibliography{supplement}